\documentclass[letterpaper,12pt]{article}   
\usepackage{osajnl2} 
\usepackage[draft]{hyperref} 
\usepackage{graphicx}
\usepackage{amsmath}

\begin{document}

\title{Feedback-Optimized Operations with Linear Ion Crystals}


\author{J. F. Eble,$^{1,*}$ S. Ulm,$^1$ P. Zahariev,$^2$ F. Schmidt-Kaler$^1$ and K. Singer$^1$}
\address{$^1$Institut f\"ur Quanteninformationsverarbeitung, Universit\"at Ulm, \\ Albert-Einstein-Allee 11, D-89069 Ulm, Germany}
\address{$^2$ Institute of Solid State Physics, Bulgarian Academy of Science\\ Tzarigradsko Chaussee Blvd. 72, 1784 Sofia,Bulgaria}

\address{$^*$Corresponding author: johannes.eble@uni-ulm.de}

\begin{abstract} We report on transport operations with  linear crystals of $^{40}$Ca$^+$ ions by applying complex electric time-dependent potentials. For their control we use the information obtained from the ions' fluorescence. We demonstrate that by means of this feedback technique, we can transport a predefined number of ions and also split and unify ion crystals. The feedback control allows for a robust scheme, compensating for experimental errors as it does not rely on a precisely known electrical modeling of the electric potentials in the ion trap beforehand. Our method allows us to generate a self-learning voltage ramp for the required process. With an experimental demonstration of a transport with  more than 99.8~\% success probability, this technique may facilitate the operation of a future ion based quantum processor. \end{abstract}

\maketitle 

\section{Introduction} Single trapped ions have been found as promising candidate system in quantum information processing. Quantum computing with up to 8 ions already has been successfully demonstrated\cite{Haeffner2005}. However, the complexity of the control of an ion crystal in an electrostatic potential rapidly increases with the number of participating ions. Therefore, it is preferable to divide an ion trap in processing and storage regions where different actions like ion loading or ion addressing with specific laser pulses are performed \cite{Haeffner2008,Blatt2008}. A highly reliable method for the shuttling of ions in micro segmented traps \cite{Schulz2007} is essential.

In current schemes, a predefined amount of ions has been shuttled between different areas of the trap, as required for a future quantum processor. The transport of single ions in a linear Paul trap and the symmetric separation of a two-ion crystal has been reported. Here a success probability exceeding 95\% has been shown \cite{Rowe2002}. Latest research in ion transport demonstrates shuttling through an X-junction\cite{Blakestad2009}. Even optimal control theory has been used to evaluate best time-dependent potential alterations for fast non adiabatic transport of ions through the trap \cite{Huber2008} and extended calculations \cite{Reichle2006} are necessary to obtain proper results. However, the theoretical outcome of optimal control can't be directly transferred to the experiment because the calculated potentials are derived from a trap model which does not necessarily exactly match the real experimental situation. Fabrication imperfections cause aberrations between the real trap geometry and the theoretical model. Patch charges on the surface of the trap electrodes may even worsen the situation because voltage changes in the order of 10~$\mu$V can lead to a completely different potential for the ions. Similar problems may also occur for neutral atoms, where a deterministic transport has been accomplished by controlling the motion of a standing-wave in a dipole trap\cite{Kuhr2001} and an optimal control scheme was proposed to improve the fidelity in a collisional gate \cite{Grangier2009}.

In our approach for controlling a multi-ion crystal our aim is to automate most of the operational building blocks. We are using the information from the observation of the ion crystal to feedback control the trap potentials in a robust way. Thus we have realized the transport of ions over 1~mm, the separation of single ions from a linear crystal and the re-joining of crystals in a realistic trap and without any prior knowledge of the potentials. Potential changes are sensed from ion locations and compensated automatically by the feedback system. Feedback techniques are commonly used with ion traps in situations such as feedback cooling of ions\cite{BUSHEV}, as well as error correction \cite{Chiaverini2004} or teleportation \cite{Barrett2004,Riebe2004}, as in all those cases the next steps of operation depend on the information read-out from the quantum system itself.

Photographic recording of a single ion was first realized in a radio frequency trap \cite{Neu1980}. For single fluorescing neutral atoms in a magneto-optical or dipole trap, the discrete levels of fluorescence prove the trapping of zero or one atom \cite{Grangier2001}. A sudden step is associated with the arrival and departure of individual trapped atoms\cite{Hu1994,Hau1996}. Sensitive CCD cameras allow for space and time resolved observation and imaging of single fluorescing atoms or ions which is essential for the work presented.

The paper is organized as follows: A description of the experimental setup as well as a short overview of the potential simulation is given in section \ref{expsetup}. Automatic loading and detection of a certain number of ions is demonstrated in section \ref{automaticloading}. In section~\ref{ionposition}, we explain our method for estimating the position of trapped ions. Positioning ions via feedback-control and displacing them to a given position is shown in section \ref{PIregulate}. We continue with the automatic separation of an ion crystal with an adaptive gain control and the splitting into separated potentials in section 6. We conclude with a discussion of future applications and improvements of our method.

\section{Experimental setup} \label{expsetup} 
\begin{figure}[htb]
	\centering
	\includegraphics[width=8 cm]{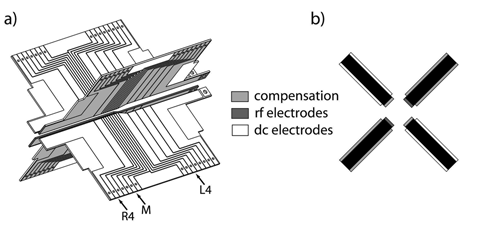}			
	\caption{a.) Sketch of the segmented linear Paul trap with dc-electrodes depicted in white and rf-electrodes in dark gray respectively. Compensation electrodes are colored light gray. b.) Front view showing that the rf-electrodes only cover two front faces of the blades. The other two are utilised as compensation electrodes.}
	\label{trap}
\end{figure}

The segmented linear Paul trap consists of four blades, each featuring a total of 15 independent dc segments \cite{Huber2008}. As the segments in the trap region are the most important, the middle electrode where the ion crystal is initially trapped is labeled as segment ~M, wheras the segments to the left and to the right of the visual focus of the camera are labeled with increasing indices as \{L1, L2,..\} and \{R1, R2,..\}, respectively. The blades are assembled in a X-shaped manner. Each blade has an additional electrode on the edge facing to the ion. Two of those opposing segments are connected to the rf~supply whereas the other two opposing are used for compensation (see Fig.~\ref{trap}).

Blade material is polyimide\footnote{Material P97, Isola AG, Germany} with a 18 $\mu$m copper plating on both sides, the strip lines are produced using standard  lithography and etching techniques. The blades are electropolished to decrease surface roughness, additionally they are coated with gold in order to become a chemically inactive surface. The trap region we use consists of eight 700~$\mu$m wide segments separated by 100~$\mu$m wide strip lines. The radial distance between two trap electrodes equals 2 mm. The trap is housed in a stainless steel vacuum chamber. The base pressure is below $10^{-12}$~mbar. The rf peak-peak voltage for the radial confinement equals $U_{rf}=400$ V$_{pp}$ at $\Omega/2\pi=13.4$~MHz resulting in a radial trappung potential with $\omega_{rad}$ = 431.65 kHz/2$\pi$, exceeding the axial confinement. This ensures that up to ten ions arrange in a linear configuration. Fast dc-voltage control for each trap electrode is accomplished via a PC by an array of digital analog converters\footnote{DAC8814, Texas Instruments}. Their voltages range from -10~V to~10~V with a resolution of $16$~bit resulting in a smallest step-size of 300~$\mu$V. Each voltage supply is low-pass filtered (cutoff frequency 390~Hz).

The ionization of $^{40}$Ca atoms is accomplished with a two-photon process by laser light near 423~nm and 374~nm \cite{Gulde2001}. For optical cooling and excitation, we illuminate the ion with laser light near 397~nm, 866~nm and 854~nm and observe continuous fluorescence. The detection system consists of a specifically designed lens with NA~=~0.30 which is placed 61~mm from the trap center at an angle perpendicular to the trap axis, and an EMCCD camera\footnote{electron multiplying charge coupled device, Andor iXon DV885} featuring 1004 x 1002 pixels with a size of 8~x~8~$\mu$m$^2$. A distance calibration with higher accuracy is obtained by measuring the axial frequency and the distance of two ions. By applying a rf voltage to segment~L6, an ion oscillation in axial direction is stimulated. The currently chosen potential yields an axial frequency of $226.3\pm 0.2$~kHz, the ions inter distance on the camera picture is $21.85\pm 0.02$~pixel. This results in a distance calibration of $0.6908\pm 0.0005\,\mu $m/pixel \cite{James1998}. The axial potential distribution may be simulated with the boundary element method using a three dimensional model of the trap \cite{SingerRMP2009}. With given voltages on the individual segments, we extract the resulting axial potential.

\section{Automatic ion loading and amount determination} \label{automaticloading} For the determination of the amount of ions in the trap, images are taken by the EMCCD camera. These images contain the count distribution $C(h,v)$ with \{h, v\} denoting the pixel position in horizontal and vertical direction, respectively. In the following, we describe a fast, real-time image analysis used to determine the number of ions and their positions. Compared to a standard off-line two dimensional Gaussian fit, our real time method allows for a fast feedback, with a slightly reduced position accuracy.

\begin{figure}[htb]
	\centering
	\includegraphics[width=8 cm]{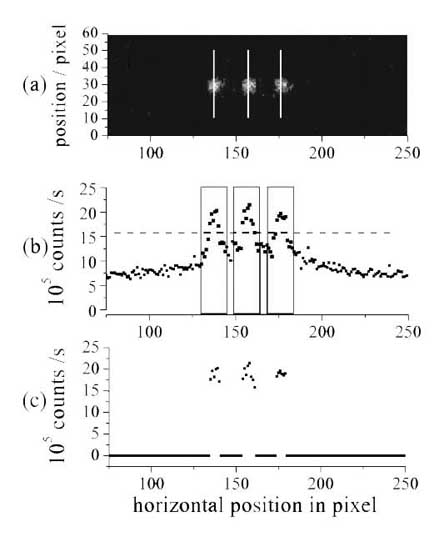}			
	\caption{Ion position determination: (a) Camera image of fluorescing ions with vertical markers to indicate the ion positions. (b) Vertically summed counts $C(h)$ with threshold parameter, here 60\%, as dashed line and boxes wherein precise ion location takes place. (c) Vertically summed counts $\tilde{C}(h)$ (values below threshold are set to zero) for amount and coarse location determination  of the ions.}
	\label{locateIon}
\end{figure}

As the ion crystal only illuminates a small area on the EMCCD chip, we choose a region of interest of  60~$\times$~250 pixel from the full image of the camera. We sum the EMCCD counts over each column $C(h)=\sum\limits_{v}C(h,v)$ (see Fig.~\ref{locateIon} (b)). To get the amount of ions, we compute the maximum of $C(h)$ and introduce a threshold parameter, which is varied between the average background noise $B$ and $C_{max}$. With this threshold parameter, it is possible to discriminate between closely spaced ions even if the fluorescence is overlapping and between unequally fluorescent ions stemming from the Gaussian profiled exciting laser beam. The background noise originates from stray light which is reflected back from the trap and readout noise of the EMCCD. In the next step, we set each value in $C(h)$ which is below the threshold parameter to zero and get the array $\tilde{C}(h)$,see Fig. \ref{locateIon} (c), containing regions with counts and regions with zeros, yielding the number of ions in the crystal. While continuously analyzing the current camera picture we load a predefined number of ions by opening and blocking the ionization light. The loading efficiency for any desired number between 1 and 10 ions is 100~\% if the loading rate and the potential shape is chosen properly.

\section{Ion position determination} \label{ionposition}

For each ion, we determine the position h$_{\text{ion}}$ by utilising the following method
\begin{equation}
  h_{\text{ion}}=\frac{\sum\limits_{h}{h\{C(h)-B\}}}{\sum\limits_{h}{\{C(h)-B\}}},
	\label{hIon}
\end{equation}
where $\{C(h)-B\}$ are pixel counts corrected by the noise level and $h$ is chosen such that it covers the range of only one ion. On a recorded data set of 3000 images with an exposure time of $\tau$=100ms, we made a comparison between our method for determining the position and the two dimensional Gaussian fit. Our method reached an accuracy of 170~nm, and the with a Gaussian fit yielded a 100~nm accuracy. Our method is preferable in situations where fast reaction time and robustness to variations of C(h) is crucial. As expected, the accuracy increases with $\sqrt{\tau}$.

\begin{figure}[htb]
	\centering
	\includegraphics[width=8 cm]{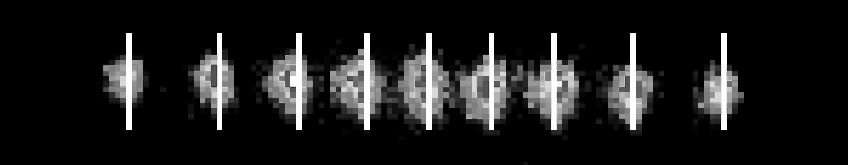}
	\caption{Ion crystal with automatically determined equilibrium positions. The exposure time $\tau$ was set to 150~ms.}
	\label{IonStringLong}
\end{figure}

In the experiment, see Fig.~\ref{IonStringLong} we apply the location algorithm to an ion cyrstal consisting of eight ions, for a proper linear arrangement the rf amplitude is increased to 650~V$_{pp}$. Assuming a harmonic potential in the axial direction, the determined locations agree within about 1/7th of one pixel with theoretical values \cite{James1998}. The automatically determined ion locations relative to the centered ion are (values in $\mu m$):
\begin{center}
	\begin{tabular}{cccccccccc}
		experimental: & -36.0 & -25.0 & -15.9 & -7.6 & 0 & 8.1 & 16.4 & 25.3 & 35.8 \\
		theoretical: & -35.78 & -25.22 & -16.28 & -8.00 & 0 & 8.00 &  16.28 & 25.22 & 35.78
	\end{tabular}
\end{center}

\section{Feedback ion position regulation} \label{PIregulate} To keep a single ion or an ion crystal fixed at one position in presence of external disturbances, we performed a feedback control. The trapping potential is created by a negative voltage on segment~M and positive voltages on segment~L1 and R1. We used a camera exposure time of 25~ms (maximum available gain). The feedback control was implemented by a digital proportional and integral (PI) controller which is fed with the position information $x_{\text{act}}$ from the ion position determination algorithm described above. Comparing the actual value with a target value $x_{\text{aim}}$, the PI control regulates the ion position in axial direction by changing the voltage $V_{L1}$ = $V_{L1}^{\text{old}}$ - $\Delta V_{L1}$ of segment~L1. The PI controlled voltage change $\Delta V_{L1}$ is calculated as
\begin{equation}
\Delta V_{L1}=\text{P}\cdot \left(x_{\text{aim}}-x_{\text{act}}\right)+\text{I}\cdot\sum\left(x_{\text{aim}}-x_{k}\right),
\label{equ:PI}
\end{equation}
where the integral term is updated in each step. It was found that a derivative term not improved the regulation.
In the test routine, the ion was regulated alternating between the initial ion position and a position shifted 60~pixels to the left, which corresponds to a distance of 41.4~$\mu$m. The optimal PI gain is found for $\text{P}=7\,\text{mV}/\text{pixel}$ and I~$=1\,\text{mV}/\text{pixel}$ but the regulation works still if this values are set in between 0.5 and 2 of the optimum. Please note that the optimal gain depends on the trapping potential, see section~\ref{split}.

\begin{figure}[htb]
	\centering			
		\includegraphics[width=10.0cm]{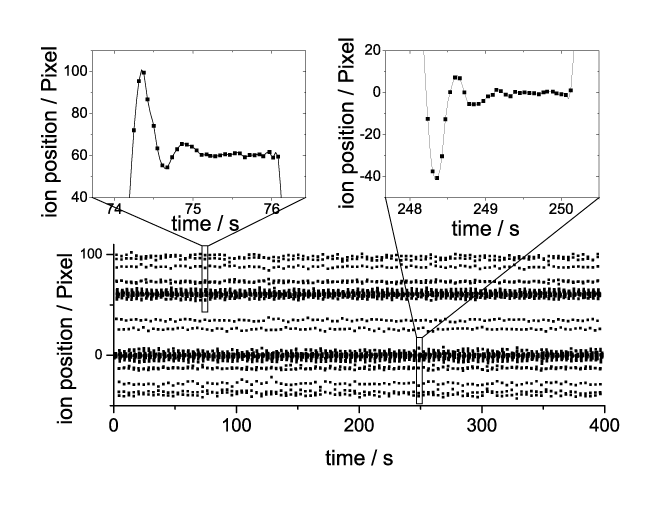}	
	\caption{Feedback regulated ion position as a function of time, consisting of 100 consecutively executed forth and back regulations over 60~pixel. Insert (a) shows a zoom of a single regulation process for moving the ion from position A to B whereas insert (b) shows the regulation for transporting the ion from B to A, respectively. Here, the regulation between A and B is accomplished within 600~ms.}
	\label{PIDlong}
\end{figure}

\begin{figure}[htb]
	\centering			
		\includegraphics[width=10.0cm]{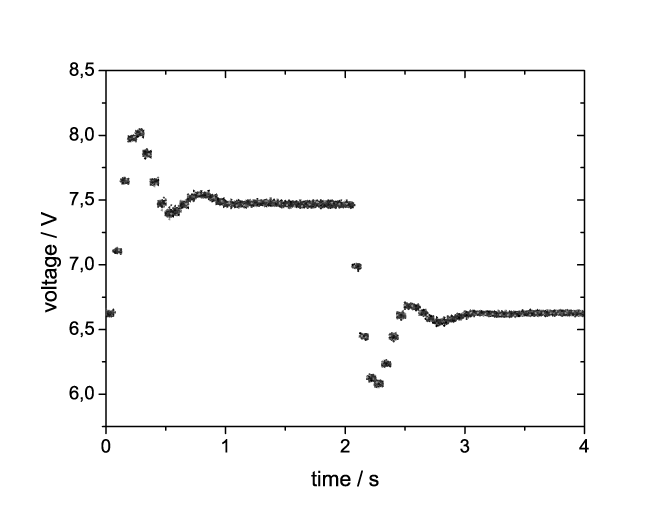}
	\caption{Control voltage alterations on segment~L1 for 100 cycles with different colors for each loop. The repetitions, lying upon each other, indicate similar reactions of the system.}
	\label{PIVolt}
\end{figure}

In Fig.~\ref{PIDlong} the position of the ion is shown during the regulation at two distinct locations A and B as a function of the time. From a large number of transports, we determine a success probability of 99.8~\% where the new position is achieved within a timespan of 600~ms. Interestingly, the required control voltage V$_{L1}$ does barely show any variation for the consecutive transports, see also Fig.~\ref{PIVolt}. The algorithm has "learned" the way how to transport an ion. Only if external disturbances occur, the PI regulation will adapt the voltage ramp. Due to the robustness of our detection algorithm, the PI controller can handle strong disturbances of the trapping potential. It is only limited by the extension of the laser beam diameter with a FWHM of 76$\mu$m, as we need a sufficient number of fluorescence photons to gain an adequate signal-to-noise ratio for the PI regulator. A typical application for this kind of PI control may be a long time ion position regulator in ion traps where patch charges or other disturbances cause an axial ion drift in the time domain of the performed experiments.

\section{Automatic splitting of an ion crystal} \label{split} The separation is investigated in two different ways: Performing a symmetric separation, the ion crystal is divided such that two equal parts move equally far apart from the initial position into well separated axial potential wells. Typically, the initial position of the crystal is exactly above one trap control segment, whose voltage is ramped down \cite{Rowe2002}. In the case of an asymmetric splitting, one or more ions may be kept at fixed positions while another part of the crystal is split off. Here, the position of the ion crystal is not limited to be exactly above a specific segment.

\begin{figure}[htb]
	\centering	
		\includegraphics[width=8.0cm]{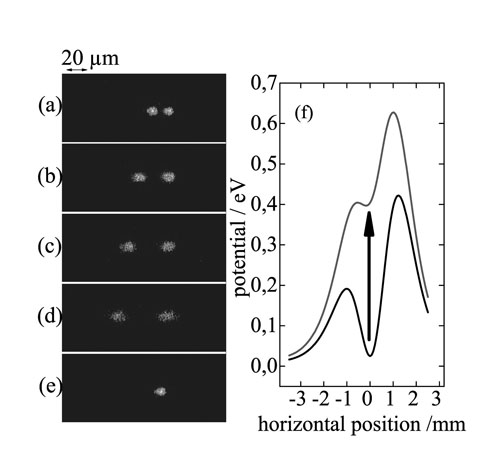}		
	\caption{Automatic asymmetric splitting of a two-ion crystal: Camera pictures of fluorescent ions - before (a), during (b)-(d) and after the splitting process (e). Pictures (b)-(d) are taken after decreasing the potential depth and moving the left hand ion. (f) Simulated potentials showing the alteration of the potential during splitting process. Starting with a deep potential at the beginning of the splitting process the voltages are changed in such a way that one ends with a shallow potential briefly before the crystal splits. In (e), only the right hand ion stays in the trapping potential. Here, the potential is set to its initial values, such that the ion location is precisely in the middle of the two ions depicted in (a).}
	\label{splitIon}
\end{figure}

To split a two-ion crystal in the asymmetric mode, we start with a deep axial trapping potential created by a negative voltage on segment~M and positive voltages on segment~L1 and segment~R1, respectively. This configuration results in a localization of the ion crystal above segment~M. We reduce the potential depth by lowering the voltage supplied to segment~M and control the axial position with segment~L1, while segment~R1 is held at a fixed value.
The lowering of the potential is performed in such a way that the inter-ion distance increases linearly. With the relation between the minimal ion-ion distance in a harmonic potential and the axial frequency $\Delta x\propto\omega_{ax}^{-2/3}$ \cite{James1998} and by using the relation $\omega_{ax}\propto {V_{ax}}^{1/2}$  between the axial trapping frequency $\omega_{ax}$ and the axial confinement voltage $V_{ax}$, we find
$\Delta x\propto{V_{ax}}^{-1/3}\;$. In a segmented linear Paul trap with ions above segment~M the axial voltage is given by : $V_{ax}=V_{L1\,\text{or}\,R1}-V_{M}$, whereas the lower lateral voltage is taken. A linear increase of the ion-ion distance can be described by $m\,t+\Delta x_0$ where $\Delta x_0$ denotes the initial distance and $m$ the voltage alteration velocity. The voltage alteration on segment~M for decreasing the potential depth is then given by:
\begin{equation}
V_{M}=V_{L1\,\text{or}\,R1}-\frac{A}{\left(m\,t+\Delta x_0\right)^3}\, ,
\label{v10alteration}
\end{equation}
where the constant $A$ is deduced from $V_M(t=0)$. During the decrease of the trap depth, the minimum of the potential is also shifted but this is balanced with control segment~L1 via the PI-control. However, the PI gain parameters need to be dynamically adapted for this task, contrary to the transport of ions in a potential with fixed axial trapping frequency. With a change of the axial trapping frequency, the system response changes accordingly. The spring constant in a harmonic potential is given by $\textbf{F}=-k\textbf{x}$ with $k=m\omega_{ax}^2$. Changing $k$-values have to be compensated with the total PI-gain G$_{PI}$ acting as a multiplication factor on $\textbf{F}$: $G_{\text{PI}}\propto k\propto \omega_{ax}^2$. With the relation $\omega_{ax}\propto\sqrt{V_{ax}}$ the gain is given by $G_{\text{PI}}\propto V_{ax} = V_{L1\,\text{or}\,R1}-V_{M}$. By multiplying this total gain with the PI-values from equation(\ref{equ:PI}), the ion positioning is achieved for altering axial trapping frequencies. When the potential is deformed, the inter-ion distance increases. If the Coulomb repulsion energy exceeds the potential depth, then ions are leaving the trap, and the desired number of ions is kept in the crystal. In the experiment, we find an ion-ion distance of 60$\left(1\right)$\,$\mu$m when only one ion is kept in the potential. The Coulomb energy reads $E_{\text{Coul}}=\frac{1}{2}\frac{e^2}{4\pi \epsilon_{0}}\frac{1}{d}$, which corresponds to a potential depth of $\Phi=E/e$ is 25$\left(1\right)$\,$\mu$V. The loss of an ion can either be detected by a reduction of the fluorescence light on the EMCCD or from a sudden jump of the position of the remaining ion(s). Thus, we may even detect a non-fluorescing ion leaving the potential making this method applicable to ion crystals consisting of mixed ion species.

\begin{figure}[htb]
	\centering	
		\includegraphics[width=8.0cm]{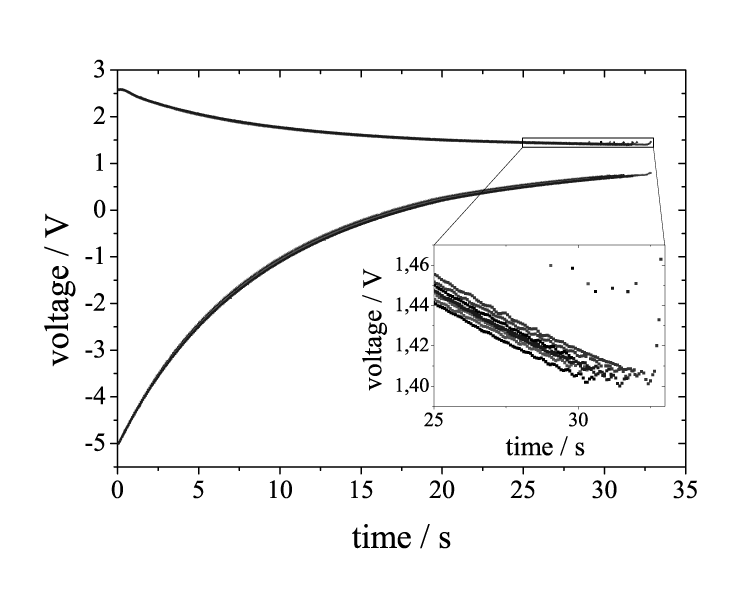}		
	\caption{Voltage alterations during repeated splitting experiments. The potential shape is manipulated with segment~M whereas segment~L1 regulates the position of the ion. Each splitting cycle is plotted with different colors. The insert shows a zoom into the control voltage alteration at the end of the separation process.}
	\label{splitIonsRepeat}
\end{figure}

To show the high degree of automation, the separation algorithm has been repeated many times, see Fig.~\ref{splitIonsRepeat}. Both, the general shape of the voltage ramps and the control electrode voltages at the point when the ion crystal is splitted differ only slightly, here about 10 mV, from shot to shot. 

\newpage
We reach a success probability of $95$~\%. We have also realized the symmetric splitting of ion crystals which were positioned above segment~M by starting with control voltages of 5\,V, -2\,V, -5\,V, -2\,V and 5\,V for segment~L2 to segment~R2. When the potential depth is reduced by changing V$_M$, the regulation of V$_{L1}$ guarantees that the center of mass is not changed.

\paragraph{Automatic separation and reunification of ion crystals:} \label{reuinfication} While in the previous section a part of the ion crystal was split off and lost, 
\begin{figure}[htb]
	\centering	
		\includegraphics[width=8.0cm]{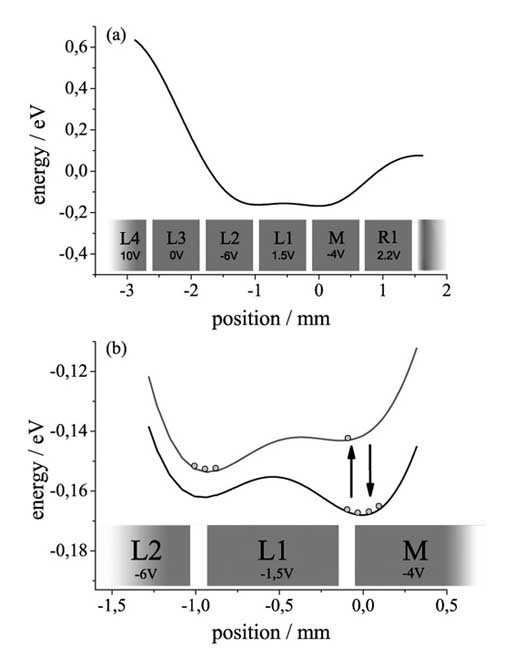}		
	\caption{Potentials during the asymmetric separation and reunification process: (a) Axial potential showing a moving minimum above the insulator between segment~L1 and L2 and a stationary minimum above the insulator between segment~L1 and M. The potential barrier on the left above segment~L3 and L4 prevents the ions from leaving the trap during the separation. The voltage configuration displays the situation in the beginning of the separation process. (b) Zoom of the axial potential showing the potential change during the separation and reunification of the ion crystal. For reasons of clarity, the upper potential has been shifted via an offset of 0.13~eV .}
	\label{splitpotential}
\end{figure}
here we will describe a protocol to (i) separate crystals into different axial potential wells and (ii) recombine the parts again into one ion crystal. 
The scheme is illustrated in Fig. \ref{splitpotential}. A confining potential is formed using the trap electrodes L4 to R1. When the voltage of electrode M is increased, the double well forms, and L1 is regulated according to the fluorescence position information. 
This way, a N-ion crystal is split in such a way that one ion is staying in its position while N-1 ions are shifted into a second potential well at the left hand side. If the protocol is reversed, the two ion crystals are recombined in the original potential.

In the experimental realization, we have used  three and four-ion crystals, and kept the position of the outermost right ion fixed while the other part of the crystal was shifted into a separated potential well. The ion crystal may be recombined by merging the potential wells. Such a separation process can also be performed without the assistance of the camera. For that, the system has to learn the right voltage alterations. Therefore one successful separation process has to be accomplished during which the voltage alterations for the segments are recorded (see Fig.\ref{splitIonsRepeat}). Higher separation velocities can be achieved by replaying the learned voltage sequence with a speed up factor of up to 20.

\section{Conclusion and Outlook} We have presented an experimental realization of self-adapting and self-regulated algorithms for the automation of fundamental transport routines in a segmented linear Paul trap which are important for quantum information processing with trapped ions. A sensitive camera for the ion detection and a software control of each trap segment is used for building a feedback loop. We show the feedback controlled positioning of an ion to specific locations in the axial direction of the trap via a software PI regulator. By creating two trapping potentials in the axial direction and merging them into a single potential well, we have shown the separation and reunification of ion strings.

For the future, we envision several improvements of our method: The detection and the overall control loop can be sped up, when we use a faster EMCCD camera or only read out a subsection of the image. In our trap, a segment width of 700$\mu$m dictates large ion-ion distances, and therefore low trap frequencies when the splitting occurs. Thus, a major increase in speed is expected when we will apply the method to ion crystals which are stored in a segmented micro ion trap with segment dimensions as small as 125$\mu$m \cite{Schulz2007}, where the trap control segments are  optimized for ion transport and splitting operations. Another improvement will be a more sophisticated feedback loop. Optimized gradient search may be helpful, especially when not only a single but multiple trap control parameters need to be adapted. Finally, we intend to apply feedback methods not only to the position of the ion crystals, but also to the internal electronic qubit states.

We acknowledge financial support by the Landesstiftung Baden-W\"urttemberg in the framework of the excellence program, the European Commission EMALI (Contract No. MRTN-CT-2006-035369) and STREP MICROTRAP (Contract No. FP6-IST-517675) and the BMBF (06UL264I). We thank W. Schnitzler for proofreading and R. Maiwald for important contributions at an earlier stage of the experiment.

\end{document}